\documentclass[a4paper,11pt]{article}
\pdfoutput=1 

\usepackage{jcappub} 
\usepackage{xcolor}
\usepackage{graphicx}
\usepackage[T1]{fontenc} 
\usepackage{isotope}
\usepackage{subcaption}
\usepackage{array}

\newcommand{\lya}{Ly$\alpha$}
\newcommand{\lyaf}{Ly$\alpha$ forest}
\newcommand{\lyb}{Ly$\beta$}
\newcommand{\lyalya}{Ly$\alpha \times$ Ly$\alpha$}
\newcommand{\lyaqso}{Ly$\alpha \times$ QSO}

\newcommand{\Hunits}{km s$^{-1}$ Mpc$^{-1}$}
\newcommand{\planck}{{\it Planck}}

\title{Baryon Acoustic Oscillations and the Hubble Constant: Past, Present and Future}
\author[a]{Andrei Cuceu,}
\author[a]{James Farr,}
\author[a]{Pablo Lemos,}
\author[a]{Andreu Font-Ribera}

\affiliation[a]{Department of Physics and Astronomy, University College London, London, UK}

\emailAdd{andrei.cuceu.14@ucl.ac.uk}
\date{\today}

\abstract{We investigate constraints on the Hubble constant ($H_0$) using Baryon Acoustic Oscillations (BAO) and baryon density measurements from Big Bang Nucleosynthesis (BBN). We start by investigating the tension between galaxy BAO measurements and those using the Lyman-$\alpha$ forest, within a Bayesian framework. Using the latest results from eBOSS DR14 we find that the probability of this tension being statistical is $\simeq6.3\%$ assuming flat $\Lambda$CDM. We measure $H_0 = 67.6\pm1.1$ \Hunits, with a weak dependence on the BBN prior used, in agreement with results from \planck\ Cosmic Microwave Background (CMB) results and in strong tension with distance ladder results. Finally, we forecast the future of BAO $+$ BBN measurements of $H_0$, using the Dark Energy Spectroscopic Instrument (DESI). We find that the choice of BBN prior will have a significant impact when considering future BAO measurements from DESI.}

\begin{document}
\maketitle
\flushbottom

\section{Introduction}






Over the last twenty years, a clear picture of the Universe has started to emerge, with Lambda Cold Dark Matter ($\Lambda$CDM)  becoming the standard cosmological model. However, with the improved precision of the latest surveys, tensions between different measurements of some parameters have also started to appear. Perhaps none have been debated more than the discrepant values of the Hubble constant, $H_0$, that measures the expansion rate of the Universe. The cosmic distance ladder has long been used to directly measure $H_0$ \cite{Riess:2009,Riess:2011,Riess:2016,Riess:2018,Riess:2019}, and the latest value from the Supernova, $H_0$, for the Equation of State of Dark energy (SH0ES) program is $H_0 = 74.03 \pm 1.42$ \Hunits \cite{Riess:2019}. On the other hand, indirect constraints using Cosmic Microwave Background (CMB) anisotropy measurements from the \planck\ satellite \cite{AstropyCollaboration:2013,PlanckCollaboration2015PlanckParameters,PlanckCollaboration:2018} give a significantly different value: $H_0 = 67.36 \pm 0.54$ \Hunits \cite{PlanckCollaboration:2018} (assuming $\Lambda$CDM).

Possible explanations for this tension are systematic errors in one or both datasets, or problems with the standard model and the need for new physics. Reanalyses of the distance ladder data (e.g. \cite{Cardona:2016,Zhang:2017,Feeney:2017,Follin:2017}) still prefer high values of $H_0$, while using most subsets of the \planck\ data yields lower values (e.g. \cite{PlanckCollaboration:2018,Bernal:2016}). The $4.4\sigma$ difference between the two $H_0$ measurements is also hard to reconcile with extensions to the standard $\Lambda$CDM model. A promising prospect is a higher value of the effective number of neutrinos, $N_{\text{eff}}$. However, the tension is only slightly reduced ($\sim3.9\sigma$), as CMB constraints rule out very high values for this parameter \cite{PlanckCollaboration:2018}.



Baryon Acoustic Oscillations (BAO) provide a standard ruler which has been evolving with the Universe since recombination. As such, probing the BAO scale at different times is a powerful tool in constraining cosmology. The best measurements of the BAO scale come from CMB anisotropy measurements at redshift $z \approx 1100$ \citep[e.g.][]{PlanckCollaboration:2018}. BAO are also present in the distribution of matter, and there are measurements at low redshifts using the clustering of galaxies \citep[e.g.][]{Beutler:2011,Ross:2015,Alam:2017}. It has also been detected in the correlation function of the \lyaf\ at $z \sim 2.4$ and in its cross-correlation with quasar positions \citep[e.g.][]{Delubac:2015,FontRibera:2014,Bautista:2017,duMasdesBourboux:2017,Blomqvist:2019,DeSainteAgathe:2019}. 

BAO data can only constrain a combination of the size of the sound horizon and the expansion rate of the Universe ($H_0$). Therefore, a constraint on $H_0$ requires extra data to calibrate the size of the sound horizon; usually CMB anisotropy measurements are used. Recently, \cite{Addison:2018} used an alternative method, introduced by \cite{Addison:2013}, that uses deuterium abundance measurements and the Big Bang Nucleosysnthesis (BBN) theory. This BAO $+$ BBN method assumes standard pre-recombination physics and gives a value of $H_0$ consistent with the \planck\ value using a flat $\Lambda$CDM model. \cite{Addison:2018} emphasized the importance of this method in providing a constraint on $H_0$ independent of CMB anisotropy measurements and the distance ladder. The focus of this work is to discuss past results of this data combination, compute the latest constraints, and investigate future implications.




The BAO measurements used by \cite{Addison:2018} come from galaxy clustering analyses \cite{Beutler:2011,Ross:2015,Alam:2017}, and the \lyaf\ \cite{Delubac:2015,FontRibera:2014}. Questions arise, however, when considering the $\sim2.5\sigma$ tension between Galaxy BAO and \lya\ BAO in the 11th and 12th data release of the Sloan Digital Sky Survey (SDSS DR11 and DR12, e.g. \cite{duMasdesBourboux:2017,Bautista:2017,Aubourg:2015}). The question of consistency between datasets, especially when it comes to combining them, has long been debated \citep[e.g.][]{Inman:1989,Charnock:2017,Nicola:2019,Adhikari:2019,Raveri:2019}. Recently, a new method was proposed by \cite{Handley:2019} to quantify tension using a new statistics they call suspiciousness.  As such, in Section \ref{sec:tension} we use this method to investigate the tension between Galaxy BAO and \lya\ BAO for the purpose of testing the reliability of their combined results. 

In Section \ref{sec:H_0} we update the constraint from BAO $+$ BBN using the latest BAO and BBN results. Compared to \cite{Addison:2018}, we add the latest BAO measurements from the Extended Baryon Oscillation Spectroscopic Survey (eBOSS) using QSO clustering and the \lyaf\ \cite{Ata:2018,DeSainteAgathe:2019,Blomqvist:2019}. We also use the latest primordial deuterium abundance results \cite{Cooke:2018}. In Section \ref{sec:DESI}, we forecast future BAO + BBN measurements of $H_0$ using the Dark Energy Spectroscopic Instrument (DESI), and discuss the role of BBN priors on future results.






\section{Galaxy BAO vs \lya\ Forest BAO}
\label{sec:tension}

\begin{figure}
\centering
\begin{subfigure}{.5\textwidth}
    \centering
    \includegraphics[width=1.0\textwidth,keepaspectratio]{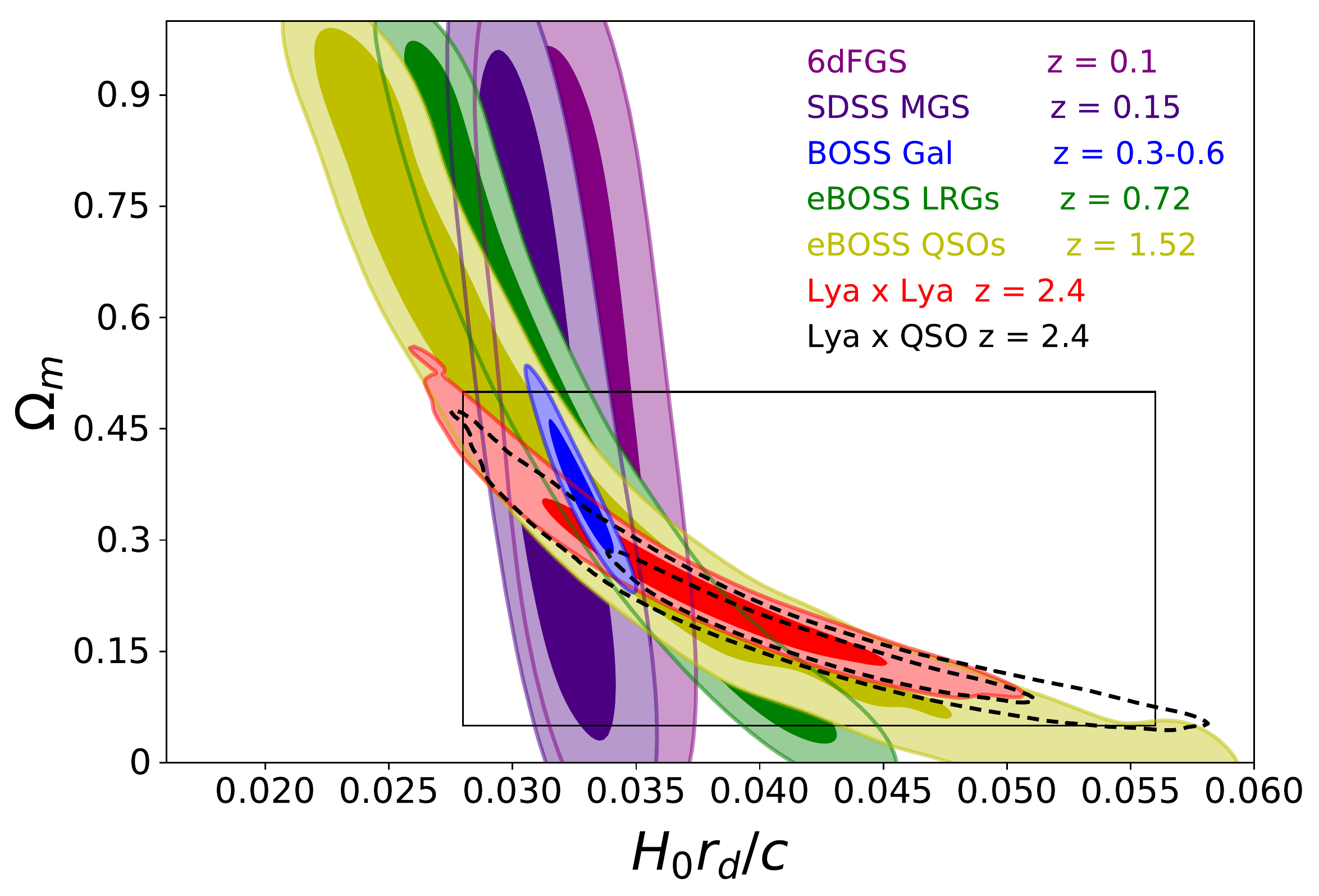}
\end{subfigure}%
\begin{subfigure}{.5\textwidth}
    \centering
    \includegraphics[width=1.0\textwidth,keepaspectratio]{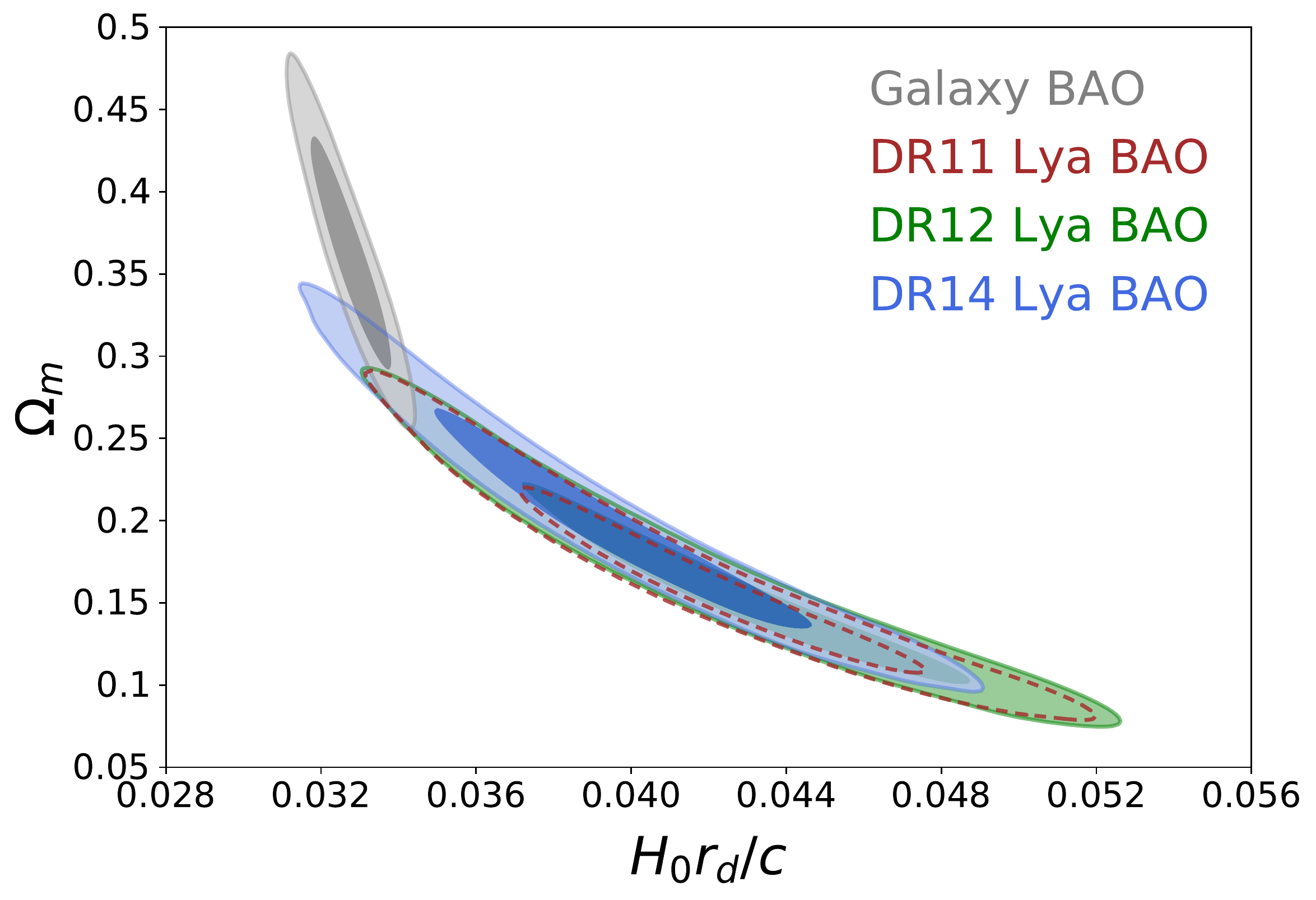}
\end{subfigure}
\caption{(Left) Parameter constraints in a flat $\Lambda$CDM cosmology from each BAO dataset individually. 
The different contour orientations are due to the different redshifts of separate datasets. The box represents the boundaries of the plot on the right with the combined BAO measurements.
(Right) Comparison of BAO constraints from galaxy clustering and different \lyaf\ measurements. The recently released eBOSS DR14 \lya\ BAO measurements are visibly more consistent with galaxy BAO than previous results from DR11 and DR12. 
This is quantified in Table \ref{tab:tension}.}
\label{fig:tension}
\end{figure}

When combining different BAO measurements, \cite{Addison:2018} split the data in two types: Galaxy BAO and \lya\ BAO, that includes both the \lya\ auto-correlation and its cross-correlation with quasars. BOSS DR11 \lya\ BAO measurements were in $\approx 2.3 \sigma$ tension with CMB predictions from the \planck\ Collaboration \cite{FontRibera:2014,duMasdesBourboux:2017}, while the samples that go into Galaxy BAO were all consistent with CMB predictions. This translated into a tension between \lya\ BAO and Galaxy BAO that can clearly be seen in the right panel of Figure \ref{fig:tension} (red dashed contours).

    
Recently, the eBOSS collaboration published the latest \lya\ BAO measurements using DR14 data \cite{DeSainteAgathe:2019,Blomqvist:2019}. They use $\sim15\%$ more quasar spectra than the previous DR12 results, and, for the first time, \lya\ absorbers in the \lyb\ region are used. With these new measurements, the tension with CMB predictions has gone down to $\sim 1.7 \sigma$. In this section we discuss the internal tensions of the latest BAO results, listed in Table \ref{tab:data}. 





\begin{center}
\begin{table}
\centering
\begin{tabular}{c c c c c}
\hline\hline            
BAO Measurement & Dataset & Reference & Tracer & $z_{\text{eff}}$ \cr 
\hline
6dFGS & 6dFGS & \cite{Beutler:2011} & galaxies & $0.106$ \cr
SDSS MGS & SDSS DR7 & \cite{Ross:2015} & galaxies & $0.15$ \cr
BOSS Gal & SDSS DR12 & \cite{Alam:2017} & galaxies & $0.38, 0.51, 0.61$ \cr
eBOSS QSO & SDSS DR14 & \cite{Ata:2018} & QSO & $1.52$ \cr
eBOSS \lyalya & SDSS DR14 & \cite{DeSainteAgathe:2019} & \lyalya & $2.34$ \cr
eBOSS \lyaqso & SDSS DR14 & \cite{Blomqvist:2019} & \lyaqso & $2.35$ \cr
\hline
\end{tabular}
\caption{Datasets measuring the BAO peak that are used in our Hubble constant analysis. We have also used other past results such as \lya\ DR11 and DR12 for our tension analysis. We assume Gaussian likelihoods for the galaxy BAO measurements, but we use the full $\chi^2$ tables provided by the \lyaf\ analyses (see \ref{app:modules}).}
\label{tab:data}
\end{table}
\end{center}

\subsection{BAO cosmology}

Studies of the BAO feature in the transverse direction provide a measurement of $D_M(z)/r_d$, while BAO studies along the line of sight measure the combination $D_H(z)/r_d = c/ H(z) r_d$, where $D_M$ is the comoving angular diameter distance, $c$ is the speed of light in vacuum, $z$ is the redshift and $r_d \equiv r_s(z_d)$ is the size of the sound horizon at the drag epoch ($z_d$).

In a flat $\Lambda$CDM cosmology, $D_M$ is given by:
\begin{equation}
    D_M(z) = c \int_{0}^{z} \frac{dz'}{H(z')}.
\end{equation}
Some of the datasets we include (6dFGS, SDSS MGS and eBOSS QSO) measure $D_V(z)/r_d$, which is a combination of the BAO peak coordinates above. $D_V(z)$ is defined as:
\begin{equation}
    D_V(z) \equiv [ z D_H(z) D_M^2(z) ]^{1/3}.
\end{equation}
The Friedmann equation in flat $\Lambda$CDM completes our model:
\begin{equation}
    \frac{H(z)^2}{H_0^2} = \Omega_{r} (1+z)^4 + \Omega_{m} (1+z)^3 + \Omega_{\Lambda},
\end{equation}
where $\Omega_{r}$, $\Omega_{m}$ and $\Omega_{\Lambda}$ are the fractional densities of radiation, matter and dark energy today (at redshift $z=0$). Furthermore, in flat $\Lambda$CDM, the dark energy fraction can be computed as: $\Omega_\Lambda = 1 - \Omega_m - \Omega_r$. In the late universe, at the redshifts probed by BAO, the radiation fraction is very small. Nevertheless, we model it assuming a fixed neutrino sector with $N_{\text{eff}} = 3.046$ and 2 massless species (the third one is massive with $m_{\nu} = 0.06$ eV and contributes to $\Omega_m$), and a CMB temperature of $T_{CMB} = 2.7255$K. This has been measured by COBE/FIRAS \cite{Fixsen:1996,Fixsen:2009}, and we consider this measurement independent of \planck. Therefore, the only free parameters in $H(z)$ are $H_0$ and $\Omega_m$.

As previously mentioned, when we measure BAO we are measuring a combination of $H_0$ and $r_d$, which means the two parameters are fully degenerate. As such, we sample their product: $H_0 r_d$. We will discuss ways to break this degeneracy in the next section, but for the purpose of investigating possible internal tensions in BAO measurements we will work in the $\Omega_m - H_0 r_d$ plane. 



\subsection{Quantifying tension}

The aim of this section is to quantify the tension between the different \lya\ BAO measurements and Galaxy BAO measurements. This tension is clear when looking at the posteriors (see right panel of Figure \ref{fig:tension}), but quantifying it is a non-trivial problem, due to the non-Gaussianity of the posteriors. There is a large number of available approaches in the literature to quantify tension between datasets, e.g. \cite{Inman:1989,Charnock:2017,Nicola:2019,Adhikari:2019,Raveri:2019}. One of the most widely used methods is the evidence ratio $R$ \cite{Marshall:2006,Trotta:2008,Verde:2013}:
\begin{equation}
R \equiv {\mathcal{Z}_{AB} \over \mathcal{Z}_A \mathcal{Z}_B},    
\end{equation}
where $\mathcal{Z}$ are evidences, $A$ and $B$ denote the two datasets on their own, and $AB$ denotes the joint results. The Bayesian evidence (the probability of the data $D$ given a model $M$: $P(D|M)$) is the normalization term in Bayes' theorem, and is usually ignored if one is only interested in the shape of the posterior. However, it has useful applications, e.g. in Bayesian Model Selection \citep[e.g.][]{Sivia:2006}, and as mentioned in quantifying concordance between datasets.

As highlighted in \cite{Handley:2019}, the $R$-statistic can hide tension when the priors are arbitrarily chosen, since it is proportional to the prior volume shared by both datasets. In this work, we will use the method introduced in \cite{Handley:2019}: We calculate the `suspiciousness' $S$ as the ratio between the evidence ratio $R$, and the information ratio $I$: $S \equiv R/I$. The information ratio is defined as: 
\begin{equation}
\log I \equiv \mathcal{D}_A + \mathcal{D}_B - \mathcal{D}_{AB},    
\end{equation}
where $\mathcal{D}$ is the Kullback-Leibler divergence \cite{kullback:1951}:

\begin{equation}
\mathcal{D} \equiv \int \mathrm{d} \theta \ \mathcal{P} (\theta) \log {\mathcal{P} (\theta) \over \pi(\theta)}, 
\end{equation}
with $\mathcal{P}$ the posterior, $\mathcal{\pi}$ the prior, and $\theta$ the parameters. 

The suspiciousness $S$ can be seen as an evidence ratio $R$ from which the dependence on prior volume has been subtracted in form of the information ratio $I$. Therefore, it preserves the qualities that make $R$ a desirable statistic for dataset comparison (such as its Bayesian interpretation and its independence in the choice of parameters), but it is no longer proportional to the prior volume, and therefore it does not hide tension when wider priors are chosen. 

As described in \cite{Handley:2019}, the suspiciousness can be calibrated using the fact that, for Gaussian posteriors, it follows a $\chi^2_d$ distribution, where $d$ is the number of parameters simultaneously constrained by the combination of the datasets. From this distribution, a tension probability $p$ of two datasets being discordant by chance can be assigned as the `p-value' of the distribution.\footnote{The remaining problem is the calculation of the number of dimensions simultaneously constrained by both datasets. This is done using the Bayesian model dimensionality (BMD) introduced in \cite{Handley2019QuantifyingComplexities}. It is worth mentioning that the BMD can be smaller or larger than the number of constrained parameters in our model if the posterior is significantly non-Gaussian.} While it is clear by looking at the right pannel of Figure \ref{fig:tension} that the posteriors are non-Gaussian in the case of \lya\ BAO, this method will give us an estimate of the tension between the datasets\footnote{ In addition, as discussed in \cite{Handley:2019}, these posteriors can be `Gaussianised' using Box-Cox transformations \cite{Schuhmann:2016}, which preserve the value of $\log S$. }. We use {\tt Polychord} \cite{Handley:2015a,Handley:2015b} to sample our posteriors and compute evidences.


We use the three \lya\ BAO measurements published by the BOSS and eBOSS collaborations using SDSS data releases 11, 12 and 14. We compare each of these with the combined Galaxy BAO sample within a flat $\Lambda$CDM cosmology, and present the tension statistics in Table \ref{tab:tension}. We compute  probability values of $\simeq1.2\%$ and $\simeq1.3\%$ for the consistency between the Galaxy BAO sample and the DR11 and DR12 \lya\ results respectively, indicating that there is a very small probability that this tension appears purely by chance. On the other hand, using the latest DR14 results we compute $p\simeq6.3\%$, consistent with the tension being statistical in nature.

\begin{center}
\begin{table}
\centering
\begin{tabular}{c|c c c c c c}
\hline\hline
datasets & $\log R$ & $\log I$ & $\log S$ & $d$ & $p(\%)$ & $\sigma$ \cr 
\hline
Gal - DR11 \lya\ & $0.35 \pm 0.19$ & $4.04 \pm 0.18$ & $-3.68 \pm 0.05$ & $2.43 \pm 0.15$ & $1.20 \pm 0.15$ & $\simeq2.5$ \cr
Gal - DR12 \lya\ & $0.26 \pm 0.19$ & $3.79 \pm 0.18$ & $-3.53 \pm 0.05$ & $2.34 \pm 0.15$ & $1.31 \pm 0.16$ & $\simeq2.5$ \cr
Gal - DR14 \lya\ & $1.93 \pm 0.19$ & $3.78 \pm 0.19$ & $-1.85 \pm 0.05$ & $2.19 \pm 0.14$ & $6.30 \pm 0.61$ & $\simeq1.9$ \cr 
\hline
\end{tabular}
\caption{Tension statistics for combining Galaxy BAO and different \lya\ BAO measurements. We show results for the R-statistic, the Bayesian information and the suspiciosness. The Bayesian model dimensionality (d) introduced by \cite{Handley2019QuantifyingComplexities} is used to compute a p-value for the suspiciosness, and we use this to compute the approximate number of standard deviations for this tension. The older DR11 and DR12 \lya\ results give small p-values indicating a small probability of this tension being statistical in nature. On the other hand, the recent DR14 results show better agreement with the Galaxy BAO results.} 
\label{tab:tension}
\end{table}
\end{center}


\section{BAO and the Hubble Constant}
\label{sec:H_0}

BAO data must be combined with other measurements in order to break the $H_0 - r_d$ degeneracy and obtain a constraint on $H_0$. The sound horizon at the drag epoch is given by: 
\begin{equation}
    r_d = \int_{z_d}^{\infty} \frac{c_s(z)}{H(z)}dz,
\end{equation}
where $c_s(z) = c[3 + \frac{9}{4} \rho_b(z)/\rho_\gamma(z)]^{-1/2}$ is the speed of sound in the baryon-photon fluid \cite{Aubourg:2015}, $\rho_b(z), \rho_\gamma(z)$ are the baryon and photon densities respectively, and $z_d$ is the redshift of the drag epoch. Precise computations of $r_d$ require a full Boltzmann code, however, following \cite{Aubourg:2015}, we use a numerically calibrated approximation to avoid the additional computational cost:
\begin{equation}
    r_d \approx \frac{55.154 \exp[-72.3(\omega_\nu + 0.0006)^2]}{\omega_m^{0.25351} \omega_b^{0.12807}} \text{ Mpc},
    \label{eq:rd_approx}
\end{equation}
where $\omega_X = \Omega_X h^2$, and $X = m, \nu, b$ are matter, neutrinos and baryons respectively, and $h=H_0/100$ with $H_0$ in [km s$^{-1}$ Mpc$^{-1}$]. This approximation is accurate to $0.021\%$ \cite{Aubourg:2015} for a fixed neutrino sector with $N_{\text{eff}} = 3.046$ and $\sum m_\nu < 0.6$ eV. Our main results are also benchmarked against independent runs using CosmoMC \cite{Lewis:2002}, which uses the Boltzmann solver CAMB \cite{Lewis:1999}, to validate the approximation.

BAO measurements provide a good constraint on $\Omega_m$, and, as discussed, the neutrino sector is fixed to the minimal mass\footnote{Small deviations from the minimal neutrino mass, within the range allowed by current CMB constraints, would not have a large impact on our results. }. Therefore, to compute $r_d$, only a measurement of the baryon density, $\Omega_b h^2$, is still needed. \planck\ results currently provide the best constraints on $\Omega_b h^2$, however, the goal of this work is to constrain $H_0$ without using CMB anisotropy information. As such, we instead use primordial deuterium abundance measurements and BBN to put a constraint on the baryon density.

\subsection{BBN measurements}
Deuterium is one the most widely used primordial elements for constraining cosmology because of its strong dependence on the baryon density \cite{Cyburt:2016}. An upper bound can easily be placed on the primordial deuterium abundance because there are no known astrophysical sources that can produce significant quantities of deuterium \cite{Epstein:1976,Prodanovic:2003}. Deuterium can, however, be destroyed, and as such a lower bound on the abundance requires finding pristine systems with the lowest possible metallicities. These systems have undergone only modest chemical evolution, so they provide the best available environments for measuring the primordial deuterium abundance (see \cite{Cyburt:2016} for a review). Recently, \cite{Cooke:2018} reported a one percent measurement of the primordial deuterium abundance using 7 near-pristine damped \lya\ systems (DLAs). However, the sample size should be greatly improved upon with the next generation of 30m telescopes \cite{Grohs:2019}. 


To obtain a constraint on $\Omega_b h^2$, the deuterium abundance must first be converted to the baryon to photon ratio, $\eta$ \cite{Cooke:2016}. The required calculations \cite{Cooke:2016} need precise measurements of the cross-sections of reactions happening in BBN (see \cite{Adelberger:2011} for a review of measurements of these reaction rates). The radiative capture of protons on deuterium to produce $\isotope[3]{He}$: $d(p,\gamma)\isotope[3]{He}$, is one reaction whose cross-section is proving difficult to determine in the energy range relevant to BBN. Current laboratory measurements have an uncertainty of $\gtrsim 7\%$, and as such theoretical estimates are mostly used as they provide about $\sim1\%$ precision \cite{Cooke:2016}. We will use both theoretical and empirical results and compare them. The best theoretical estimates of the $d(p,\gamma)\isotope[3]{He}$ reaction rate come from \cite{Marcucci:2016}, and lead \cite{Cooke:2018} to compute:
\begin{equation}
    100 \Omega_b h^2 = 2.166 \pm 0.015 \pm 0.011 \qquad \text{(BBN theoretical)},
\end{equation}
where the first error comes from the deuterium abundance measurement, and the second from the BBN calculations. Using the empirical value for the reaction rate computed by \cite{Adelberger:2011}, the baryon density is:
\begin{equation}
    100 \Omega_b h^2 = 2.235 \pm 0.016 \pm 0.033 \qquad \text{(BBN empirical)}.
\end{equation}
These two results are in mild $\sim1.7\sigma$ tension with each other, but more importantly, the first measurement (using the theoretical rate) is in $\sim2.9\sigma$ tension with the latest CMB results from the \planck\ Collaboration\footnote{We use the results from \planck\ 2018 TT,TE,EE + lowE + lensing likelihoods}:
\begin{equation}
    100 \Omega_b h^2 = 2.237 \pm 0.015 \qquad \text{(\planck\ 2018\footnote{\cite{PlanckCollaboration:2018}})}.
\end{equation}

There are some prospects for solving this tension by allowing the effective number of neutrinos $N_{\text{eff}}$ to vary (see Figure 7 of \cite{Cooke:2018}). A slightly larger value of $N_{\text{eff}}$ would reconcile BBN and CMB measurements of $\Omega_b h^2$ \cite{Cooke:2018}. However, for the purposes of the present work, we use both values $\Omega_b h^2$ from BBN with the standard $N_{\text{eff}} = 3.046$ in order to study the impact of this tension on $H_0$ measurements.

\begin{center}
\begin{table}
\centering
\begin{tabular}{m{3.1cm} m{2.0cm} m{2.5cm} m{2.3cm} m{3.2cm}}
\hline\hline               
Datasets & $\Omega_b h^2$ prior & $\Omega_m$ & $r_d$ [Mpc] & $H_0$ [\Hunits] \cr
\hline
DR14 BAO $+$ BBN & theoretical & $0.302^{+0.017}_{-0.020}$ & $149.0\pm 3.2$ & $67.6\pm 1.1$ \cr
DR14 BAO $+$ BBN & empirical & $0.300\pm 0.018$ & $148.0\pm 3.1$ & $68.1\pm 1.1$ \cr
DR12 BAO $+$ BBN & empirical & $0.290\pm 0.018$ & $150.0\pm 3.5$ & $67.5\pm 1.2$ \cr
DR11 BAO $+$ BBN & empirical & $0.289^{+0.016}_{-0.021}$ & $150.3^{+3.7}_{-3.3}$ & $67.4\pm 1.2$ \cr
\hline
\planck\ 2018 & - & $0.3153 \pm 0.0073$ & $147.09 \pm 0.26$ & $67.4 \pm 0.5$ \cr 
SH0ES & - & - & - & $74.0 \pm 1.4$ \cr 
\hline
\end{tabular}
\caption{Latest DR14 BAO + BBN constraints using either theoretical or empirical $d(p,\gamma)\isotope[3]{He}$ reaction rate. We add results using the \lya\ DR11 and DR12 measurements to show the consistency in $H_0$ results. Results from the \planck\ Collaboration \cite{PlanckCollaboration:2018} and the SH0ES collaboration \cite{Riess:2019} are included for comparison. }
\label{tab:results}
\end{table}
\end{center}
\subsection{Results}

\begin{figure}
\centering
\begin{subfigure}{.5\textwidth}
    \centering
    \includegraphics[width=1.0\textwidth,keepaspectratio]{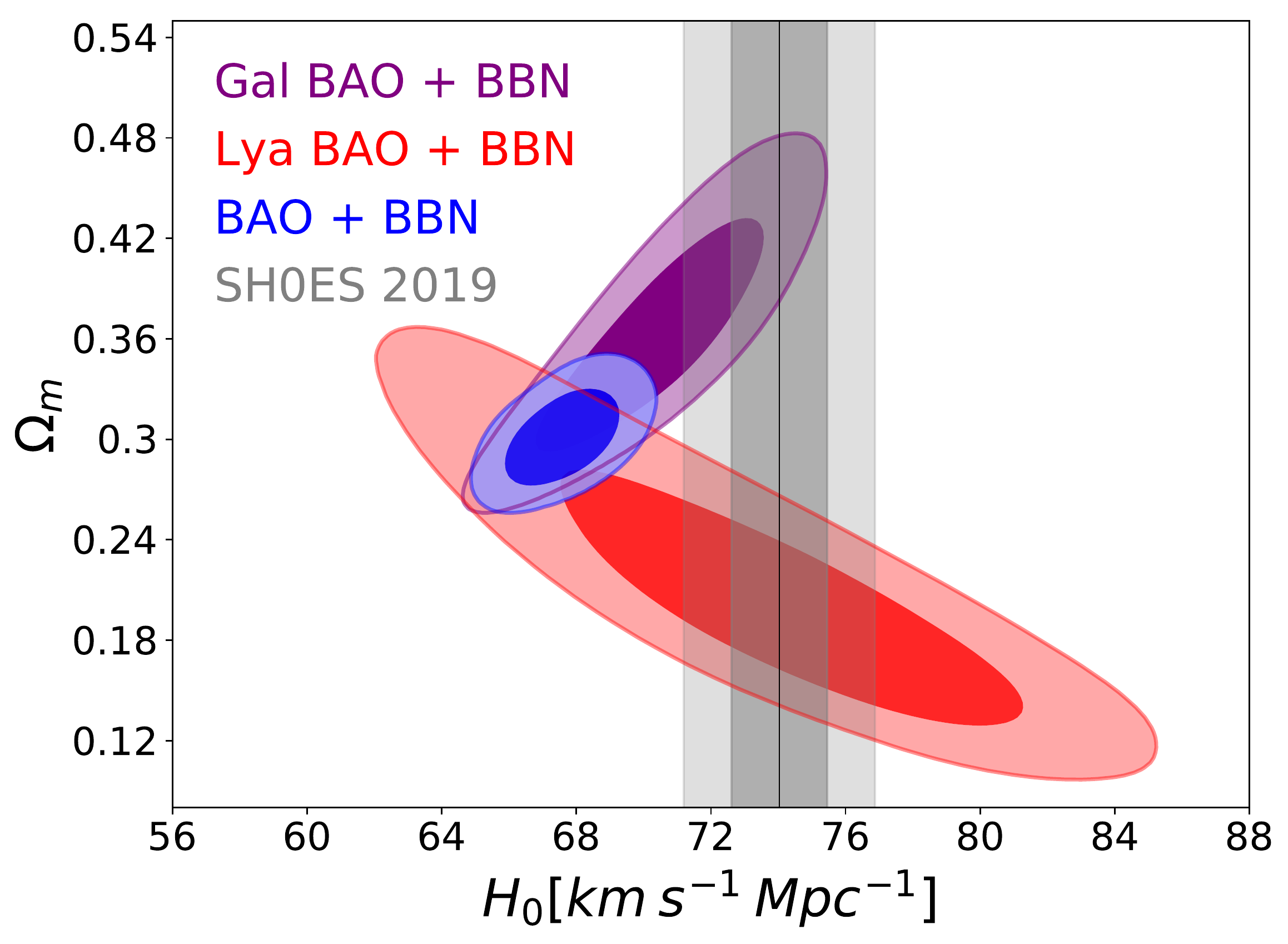}
\end{subfigure}%
\begin{subfigure}{.5\textwidth}
    \centering
    \includegraphics[width=1.0\textwidth,keepaspectratio]{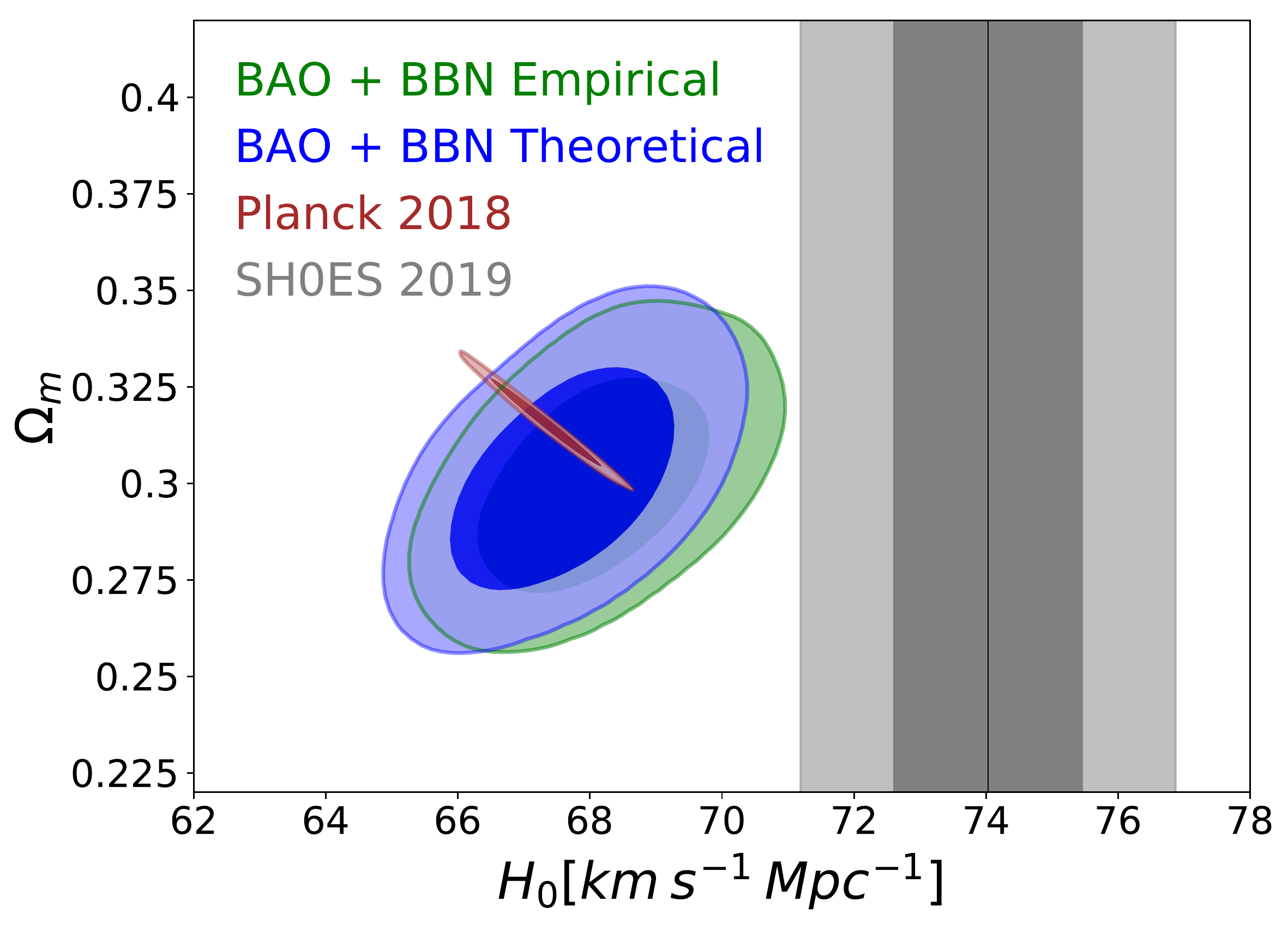}
\end{subfigure}
\caption{(Left) Current state of the art results for $H_0$ versus $\Omega_m$, independent of CMB anisotropy data. BAO data was combined with a prior on $\Omega_b h^2$ from BBN deuterium measurements (using the theoretical reaction rate). (Right) Our main results using all the BAO samples in Table \ref{tab:data}, combined with BBN using both reaction rates.}
\label{fig:H0}
\end{figure}

We combine the BAO data presented in Section \ref{sec:tension} with the two $\Omega_b h^2$ measurements from BBN deuterium abundance. Using equation \ref{eq:rd_approx}, we compute the size of the sound horizon at the drag epoch $r_d$ and obtain constraints on $H_0$. The left panel of Figure \ref{fig:H0} shows results using \lya\ BAO $+$ BBN and Gal BAO $+$ BBN, as well as their combination. Individually they are both consistent with higher values of $H_0$ (latest SH0ES results are also plotted), however once we combine \lya\ and Gal BAO, the joint constraint prefers lower, \planck-like values of the Hubble constant.

Our results using both the theoretical and empirical $d(p,\gamma)\isotope[3]{He}$ reaction rates are shown in Table \ref{tab:results} and in the right pabel of Figure \ref{fig:H0}, together with \planck\ 2018 CMB results \cite{PlanckCollaboration:2018} and the SH0ES $H_0$ measurement from the distance ladder \cite{Riess:2019} for comparison. We also add results using past \lya\ measurements (DR11 and DR12) to show the consistency in $H_0$ constraints. Both our $H_0$ measurements are consistent with the results of the \planck\ Collaboration. On the other hand, we find that our Hubble constant measurements are in strong tension with local distance ladder results of $H_0$ from the SH0ES Collaboration. Our results are in approximately $\sim3.6\sigma$ tension using the theoretical $d(p,\gamma)\isotope[3]{He}$ reaction rate, and $\sim3.3\sigma$ tension using the empirical $d(p,\gamma)\isotope[3]{He}$ reaction rate. 

An interesting result can be obtained by reframing this tension in terms of primoridal deuterium abundance. If we assume the $H_0$ constraint from SH0ES \cite{Riess:2019} is true, and we combine it with BAO data, we obtain a constraint on the baryon density of $\Omega_b h^2 = 0.0310 \pm 0.003$. Using BBN \cite{Cooke:2016}, we obtain a value for the primordial deuterium abundance of $10^5(D/H)_P = 1.38 \pm 0.25$ (this assumes $\Lambda$CDM and standard BBN). This value is $\sim4.5\sigma$ below that measured by \cite{Cooke:2018}, and well below the value derived from the interstellar medium of the Milky Way \cite{Linsky:2006}. As we discussed, there are currently no known astrophysical sources that can produce significant quantities of deuterium \cite{Epstein:1976,Prodanovic:2003}. This means D/H measurements have a robust lower limit which renders such a low value of the primordial deuterium abundance very unlikely.



We find that the relatively large difference between the two $\Omega_b h^2$ measurements from BBN has a small impact on the Hubble constant measurement from current BAO measurements, causing a shift on the best fit value of $H_0$ of about $\sim0.5\sigma$. However, with improving BAO data from the next generation of LSS experiments such as DESI \cite{Aghamousa:2016} or Euclid \cite{Laureijs:2011}, this might change. In the next section, we investigate the advances that DESI data will allow in measuring the Hubble constant independent of CMB data, and the potential impact of BBN tensions on future results.

\section{Implications for DESI}
\label{sec:DESI}

The next generation of LSS experiments will be spearheaded by the Dark Energy Spectroscopic Instrument (DESI), starting in 2020. This spectroscopic galaxy survey will cover 14000 square degrees, and measure BAO using both galaxy clustering and the \lyaf\ \cite{Aghamousa:2016}. It will target Luminous Red Galaxies (LRGs) at redshifts $0.4 < z < 1.0$, Emission Line Galaxies (ELGs) at redshifts $0.6 < z < 1.6$, quasars at redshifts $0.6 < z < 2.1$ for clustering only, and quasars at redshifts $2.1 < z < 3.5$ for both clustering and \lyaf\ measurements \cite{Aghamousa:2016}. DESI will also target bright galaxies at redshifts $0 < z < 0.5$ in order to take advantage of the times when moonlight prevents efficient observation of faint targets. This wide redshift coverage means that DESI will be able to precisely constrain the evolution of the Universe up to redshift $\sim 3.5$. Forecasts for future $H_0$ constraints from DESI combined with baryon density measurements from the CMB were presented in \cite{Wang:2017}. Our objective in this section is to forecast future DESI BAO $+$ BBN constraints on the Hubble constant, and to discuss the role of the discrepant values of the $d(p,\gamma)\isotope[3]{He}$ reaction rate.



\begin{figure}
\centering
\begin{subfigure}{.5\textwidth}
    \centering
    \includegraphics[width=1\textwidth,keepaspectratio]{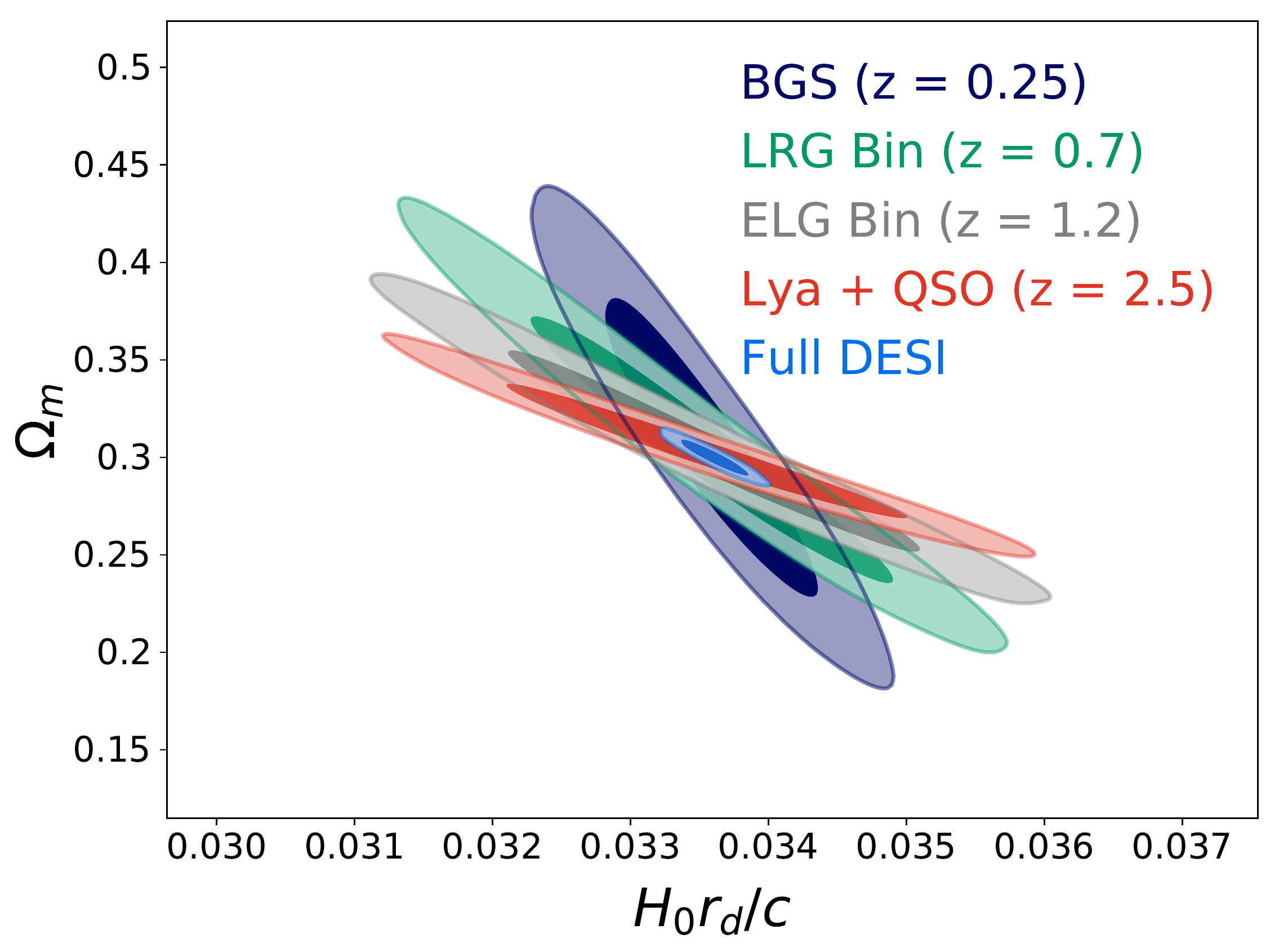}
\end{subfigure}%
\begin{subfigure}{.5\textwidth}
    \centering
    \includegraphics[width=1\textwidth,keepaspectratio]{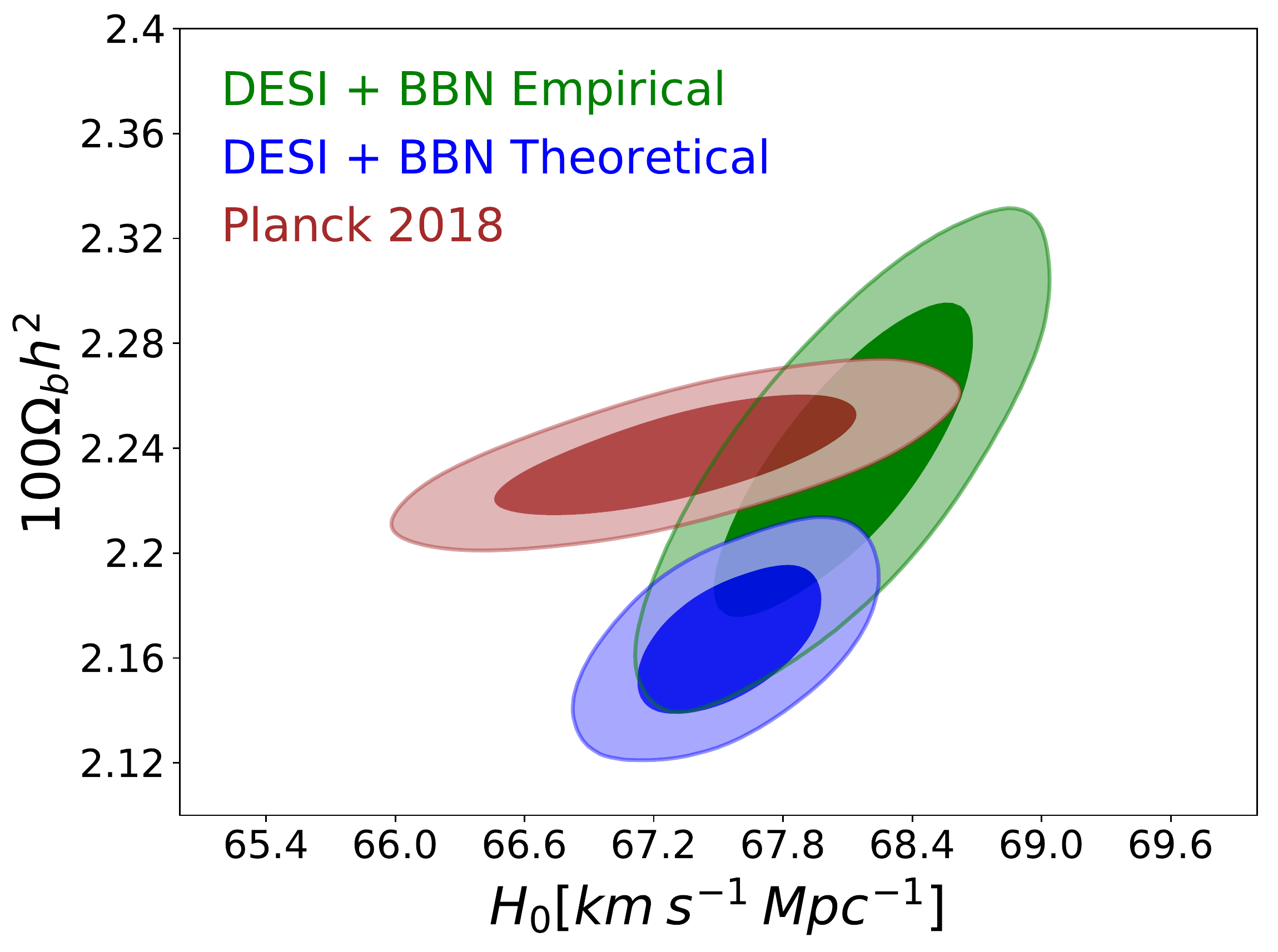}
\end{subfigure}
\caption{(left) Forecast for future BAO results within flat $\Lambda$CDM using different components of DESI. (right) Forecast for Hubble constant results using the full DESI results combined with the two $\Omega_b h^2$ priors from BBN, and the \planck\ 2018 results \cite{PlanckCollaboration:2018} for comparison. The tension in the baryon density between the BBN theoretical constraint (in blue) and the CMB (in red) can clearly be seen in this plot. This shows the importance of solving the BBN tension for the future of BAO + BBN $H_0$ measurements.
}
\label{fig:DESI}
\end{figure}

In order to study the impact of BBN tensions on future BAO + BBN measurements of the Hubble constant, we perform a forecast of the future DESI results using the uncertainties presented by \cite{Aghamousa:2016}. For our fiducial cosmology we use the BAO $+$ BBN empirical results from Section \ref{sec:H_0}. We plot results using different components of DESI as well as the combined results in the left panel of Figure \ref{fig:DESI}. For illustration purposes we only plot one LRG bin at $z = 0.7$ and one ELG bin at $z = 1.2$. With the big improvement in BAO measurements at each redshift, DESI also has the potential to give rise to inconsistent results. If this happens, it will provide a big opportunity to discover unaccounted systematic errors, unforeseen problems with our methods or potentially new physics.


Finally, we turn our attention to the main goal of this section: to quantify the impact of discrepant BBN measurements on future Hubble constant results from BAO + BBN. We perform the forecast described above for DESI and combine it with $\Omega_b h^2$ measurements from BBN using both the theoretical and empirical $d(p,\gamma)\isotope[3]{He}$ reaction rates to measure $H_0$. We plot the results in the right panel of Figure \ref{fig:DESI}. The two $H_0$ constraints are $\sim 1.2 \sigma$ apart. This means that solving the BBN $\Omega_b h^2$ discrepancy will play an important role in next generation measurements of $H_0$ using BAO + BBN. There is hope of better laboratory measurements of the $d(p,\gamma)\isotope[3]{He}$ reaction rate from the Laboratory for Underground Nuclear Astrophysics (LUNA \cite{Gustavino:2014,Kochanek:2016}).




\section{Conclusions}

We use the suspiciousness statistic proposed by \cite{Handley:2019} to investigate the tension between galaxy BAO and the different \lya\ BAO measurements. When using the DR11 and DR12 \lya\ results, we find probabilities of $\simeq1.2\%$ and $\simeq1.3\%$ for the tension being statistical in nature. On the other hand, the DR14 results show better agreement, with probability of $\simeq6.3\%$.




We put an independent constraint on $H_0$ using BAO results with the sound horizon calibrated by baryon density measurements from BBN deuterium abundance studies. One of the BBN reaction rates has very poor laboratory constraints, so we have to rely on either theoretical or empirical estimates \cite{Cooke:2016,Cooke:2018}. We obtain two $H_0$ constraints: $H_0 = 67.6 \pm 1.1$ \Hunits\ using the theoretical reaction rate and $H_0 = 68.1 \pm 1.1$ \Hunits\ using the empirical one. These results are consistent with each other and with CMB results as can be seen in Figure \ref{fig:H0}. They are also consistent with past BAO $+$ BBN results \cite{Aubourg:2015,Addison:2018}, showing that the tension in DR11 and DR12 did not have a large impact on the $H_0$ constraint. However, they are both in strong ($>3\sigma$) tension with $H_0$ results from the distance ladder. Our results again highlight that the tension is not caused by systematic errors in the \planck\ analysis. 



Starting in 2020, DESI will accurately measure BAO over a wide redshift range. We use the two BBN $\Omega_b h^2$ measurements and forecast future DESI BAO $+$ BBN results. As can be seen in the right panel of figure \ref{fig:DESI}, the choice of BBN reaction rate estimate will have a significant impact on the $H_0$ constraints. Improved measurements of the $d(p,\gamma)\isotope[3]{He}$ reaction rate (e.g. from LUNA) will be required in order to obtain accurate constraints of the Hubble constant using BAO $+$ BBN.

\section*{Acknowledgements}
We would like to thank An\v{z}e Slosar, Graeme Addison and Will Handley for helpful discussions and comments on the draft, and Ryan Cooke for comments and help with BBN calculations. AC and JF were supported by Science and Technology Facilities Council (STFC) studentships. AFR was supported by an STFC Ernest Rutherford Fellowship, grant reference ST/N003853/1, and by STFC Consolidated Grant number ST/R000476/1. PL was supported by an STFC Consolidated Grant. This work used the GetDist library\footnote{https://github.com/cmbant/getdist}.

\bibliographystyle{JHEP.bst}
\bibliography{main}

\appendix

\section{\lya\ BAO Modules} 
\label{app:modules}


For \lyaf\  datasets we use the provided $\chi^2$ tables\footnote{https://github.com/igmhub/picca/tree/master/data}. These tables give the value of the $\chi^2$ as a function of the two BAO peak coordinates scaled using a fiducial cosmology:
\begin{equation}
    \alpha_\perp = \frac{[D_M(z_{\text{eff}})/r_d]}{[D_M(z_{\text{eff}})/r_d]_{fid}} \quad and \quad
    \alpha_\parallel = \frac{[D_H(z_{\text{eff}})/r_d]}{[D_H(z_{\text{eff}})/r_d]_{fid}}.
\end{equation}
We use these tables to interpolate the value of the $\chi^2$ at the points we need for our analysis. For all the other measurements we use Gaussian likelihoods with the measured means and standard deviations (6x6 covariance matrix for BOSS). We used the Astropy\footnote{http://www.astropy.org} package \cite{AstropyCollaboration:2013,PriceWhelan:2018} for the theoretical modelling of the BAO peak coordinates. 

Methods to interpolate the $\chi^2$ tables are now available as part of the popular MCMC packages CosmoMC \cite{Lewis:2002} and MontePython \cite{Audren:2012}. As such, \lya\ BAO results can now be easily included by the community in cosmological analyses.


\end{document}